\documentclass[sigconf, screen]{acmart}

\usepackage{hyperref}

\setcopyright{none}
\renewcommand\footnotetextcopyrightpermission[1]{}
\AtBeginDocument{%
  }

\begin{document}

\title{COSMIC 1001: Engaging Future Speculation on Space Exploration with Generative AI}

\author{Lingyu Peng}
\email{lingyupeng6@163.com}
\orcid{0009-0000-5964-3340}
\affiliation{%
  \institution{Future Design School, Harbin Institute of Technology, Shenzhen}
  \city{Shenzhen}
  \state{Guangdong}
  \country{China}
}

\author{Yu Liang}
\email{2024314364@stu.hit.edu.cn}
\orcid{0009-0002-9133-9554}
\affiliation{%
  \institution{Future Design School, Harbin Institute of Technology, Shenzhen}
  \city{Shenzhen}
  \state{Guangdong}
  \country{China}
}

\author{Ying Zhang}
\email{yingzhang7@zju.edu.cn}
\orcid{0009-0001-9027-7008}
\affiliation{%
  \institution{Zhejiang University, School of Software, Hangzhou}
  \city{Hangzhou}
  \state{Zhejiang}
  \country{China}
}

\author{Chang Ge}
\email{changgecn@gmail.com}
\orcid{0009-0008-9731-5258}
\affiliation{%
  \institution{Future Design School, Harbin Institute of Technology, Shenzhen}
  \city{Shenzhen}
  \state{Guangdong}
  \country{China}
}

\author{Qingchuan Li}
\authornote{Qingchuan Li is the corresponding author.}
\email{liqingchuan@hit.edu.cn}
\orcid{0000-0001-9915-2589}
\affiliation{%
  \institution{Future Design School, Harbin Institute of Technology, Shenzhen}  
  \city{Shenzhen}
  \state{Guangdong}
  \country{China}
}

\renewcommand{\shortauthors}{Peng et al.}

\begin{abstract}
Cosmic 1001 is an interactive installation that transforms space exploration history into a speculative news experience. Participants first browse a news-based archive of major space events, then pose future-oriented questions or specify conditions such as year, celestial body, or mission name. In response, AI generates a future news item including a headline, article, narration, and visual media. These outputs are accumulated in the Future Tunnel, a shared visualization where individual stories form a collective landscape of possible futures. By combining historical space events with science fiction references, the installation explores a space between documentation and imagination, treating the future not as a fixed prediction but as a visible and discussable speculation.
\end{abstract}



\begin{teaserfigure}
  \centering
  \includegraphics[width=\textwidth]{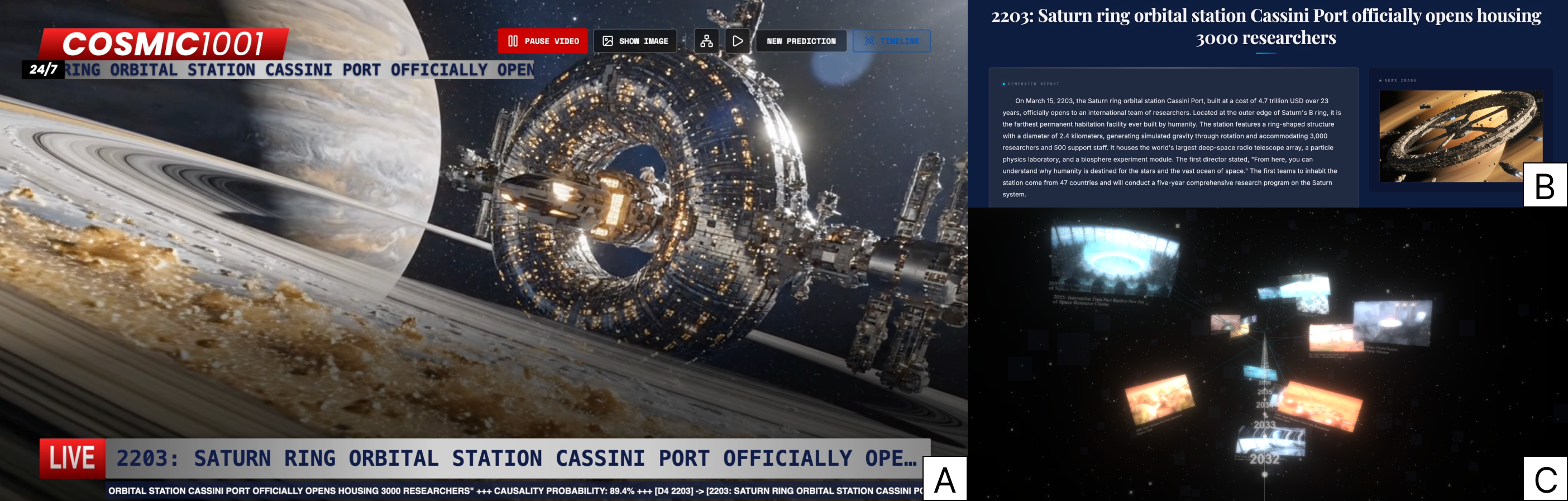}
  \caption{Three interfaces of the COSMIC 1001 installation: (A) future news broadcast, (B) future news article view with accompanying visuals, and (C) a tunnel visualization of generated news in a collective timeline of imagined futures.}
  \Description{teaser}
  \label{fig:teaser}
\end{teaserfigure}

\received{20 February 2007}
\received[revised]{12 March 2009}
\received[accepted]{5 June 2009}

\maketitle

\section{Introduction}

Public understandings of future exploration are shaped not only by technological achievements themselves, but also by the media forms through which these achievements are narrated, circulated, and experienced \cite{jasanoff2015future}. Cosmic 1001 is an interactive installation that proposes a news-based framework for engaging speculative futures of space exploration. By combining archival browsing, question driven interaction, and multimodal AI generation, it transforms space exploration from a static body of historical records into a participatory media experience. Within this framework, historical news functions not only as documentation of past events, but also as a trigger for future-oriented inquiry, while AI generated outputs turn possible futures into receivable and discussable news objects. In this way, the installation reconfigures the relationship between archive, generation, and participation, shifting users from readers of past events to participants in the construction of possible futures.

Within Cosmic 1001, news operates as an interface that connects archival media, generative AI, and public participation. It serves as a temporal scaffold for moving from historical records to speculative futures, allowing future space exploration to be received, discussed, and collectively accumulated as a shared media experience. 
\noindent\textbf{Project video:} \url{https://vimeo.com/1185854585?share=copy&fl=sv&fe=ci}

\section{System Design}
Cosmic 1001 is organized into two phases: historical news navigation and future news speculation. The first phase introduces archival records of human space exploration events through a news interface; the second phase invites users to generate speculative future news through structured prompts and open-ended questions.

\subsection{Retrospective: Navigating Historical Space Exploration Events}
In interactive installations, appropriate triggers can move viewers from passive observation toward more active engagement \cite{simon2016triggers}. When a knowledge gap becomes perceptible, this engagement may further develop into curiosity and concern \cite{veggi2024creative, loewenstein1994psychology}. Phase 1 consists of two coordinated interfaces: a news feed for detailed browsing and a timeline for chronological overview.

The waterfall-style news feed functions as the primary browsing surface of Phase 1. Events are organized by year, target celestial body, and mission type, allowing users to explore major milestones in human space exploration through headlines, summaries, and visual materials. Selecting an item opens a detailed view containing a full news-style article, representative imagery, and links to the original mission source. The content of this view is developed from an open-source chronological dataset of space exploration events\footnote{Sharmadhiraj, \textit{Timeline of Space Exploration} (JSON dataset), available via GitHub, \url{https://github.com/sharmadhiraj/free-json-datasets/blob/master/docs/science-technology/timeline_of_space_exploration.json}
, last accessed April 23, 2026.} and further supplemented with manually retrieved mission reports from official sources such as NASA Science\footnote{NASA Science, \url{https://science.nasa.gov/}
, last accessed April 23, 2026.}. These materials are then adapted into news-style text and paired with representative images and source links.

The timeline interface condenses this content into a bottom-up chronological overview, combining key dates and headlines with a background composed of the main images from the news items. It helps users situate individual missions within a broader historical trajectory and prepares the transition from retrospective browsing to future-oriented speculation.

Overall, Phase 1 turns historical browsing into a form of inquiry. Through this media framing, space exploration is presented as a continuous developmental trajectory, making historical grounding the cognitive entry point for future imagination.

\subsection{Speculation: Generating and Visualizing Future Narratives}
In Phase 2, users move from historical browsing to a generative interface for receiving future news. To support this transition, the interface provides a semi-structured input form that allows users to optionally specify a year, target celestial body, or mission name, or directly ask an open question about the future of space exploration. This input design turns user questioning into a form of scenario specification, allowing participants to define the conditions under which future news is generated.

Each speculation is presented as a multimodal future news object, including a headline, article text, voice narration, images or video, and a percentage-based realizability indicator. Drawing on visual elements of conventional news broadcasting and online news interfaces, the system presents each result as a relatively complete future news object. The realizability indicator provides an additional interpretive layer, helping users read each story as situated between speculative imagination and possible projection.

The AI generation module utilizes a Retrieval-Augmented Generation (RAG) framework to bridge historical reality with speculative narratives \cite{lewis2020retrieval}. It draws on two primary resources: first, the structured archive of space exploration events dataset introduced in Phase 1, which provides factual anchors; second, a large-scale science fiction corpus of 278,973 instruction-response pairs, used as a stylistic resource for speculative generation and described in Appendix A. In the generation pipeline, the system retrieves 218 representative sci-fi fragments from this corpus as stylistic references. By injecting these fragments into the prompt context, the system generates future news that preserves the professional tone of broadcast journalism while incorporating the vocabulary and speculative texture of science fiction.
%The AI generation module draws on two integrated datasets: a structured archive of 145 major historical events and their news descriptions, and a science-fiction corpus used as a stylistic reference. The former provides factual anchors such as missions, target celestial bodies, organizations, and technical contexts, while the latter contributes speculative vocabulary and narrative texture. In the current implementation, 218 sci-fi fragments were manually selected from a larger corpus of 278,973 instruction-response pairs and grouped into 8 thematic categories. This combination allows the system to generate future news scenarios grounded in recognizable spaceflight contexts while preserving speculative narrative variation.

Finally, all future news generated is stored within a shared visualization space, the “Future Tunnel.” Arranged along a temporal axis, it displays news created by different users and maps realizability through distance from the central axis, with entries closer to the axis treated as more likely to be realized. All viewers can browse previously received news. In this way, the tunnel transforms individual generations into a continuously expanding future archive and moves speculative news into a collectively visible space of possible futures, through which imagined futures become shared and socially meaningful \cite{jasanoff2015future}.

\section{Conclusion}
Cosmic 1001 presents an interactive installation that connects archival records of space exploration with speculative future news. By combining historical materials, news-based narration, multimodal AI generation, and shared visualization, the project guides users from browsing past missions to encountering possible future scenarios. In doing so, it proposes a participatory framework for experiencing space futures not only as abstract predictions but as publicly mediated and collectively visible narratives. More broadly, the work suggests how generative media can be used in interactive art to support shared reflection on technological
futures

\begin{acks}
This study was funded by the Guangdong Featured Innovation Project in Higher Education (Grant No. 2025WTSCX115).
\end{acks}

\bibliographystyle{ACM-Reference-Format}
\bibliography{main}


\appendix
\section{Construction of the Science Fiction Stylistic Reference Corpus}
\label{appendix:dataset}

To support stylistically consistent and narratively rich future news generation, we constructed a science fiction stylistic reference corpus for the RAG-based generation module. This corpus was not used as a supervised fine-tuning dataset. Instead, it served as an upstream retrieval pool from which representative fragments were selected and injected into the prompt context during generation.

\subsection{Corpus Provenance}
The upstream corpus contained \textbf{278,973 text items} compiled from publicly accessible science fiction resources in both English and Chinese. Source materials included public-domain novels, community-authored speculative narratives, screenwriting materials, and Chinese-language science fiction corpora. The source selection aimed to cover multiple speculative sub-genres, including hard science fiction, cyberpunk, and space opera, so as to provide stylistic diversity while maintaining a strong orientation toward technological futures.

The major source groups included: (1) classic science fiction literature from public-domain repositories; (2) community-authored prompt-to-story datasets; (3) science fiction screenplays and dialogue-based materials; and (4) Chinese-language science fiction corpora and web fiction resources. Detailed statistics of the major source groups are shown in Table~\ref{tab:dataset_stats}.

\begin{table}[H]
\centering
\footnotesize
\caption{Distribution of Major Source Groups in the Science Fiction Corpus}
\label{tab:dataset_stats}
\setlength{\tabcolsep}{4pt}
\renewcommand{\arraystretch}{1}
\begin{tabular}{p{0.56\linewidth} c c}
\toprule
\textbf{Source Group} & \textbf{Count} & \textbf{Percentage} \\
\midrule
Community-authored speculative narratives & 272,088 & 97.53\% \\
Classic science fiction literature & 3,745 & 1.34\% \\
Chinese-language sci-fi and web fiction & 2,540 & 0.91\% \\
Screenplays and other materials & 600 & 0.22\% \\
\midrule
\textbf{Total} & \textbf{278,973} & \textbf{100.00\%} \\
\bottomrule
\end{tabular}
\end{table}
\subsection{Preprocessing and Reference Library Construction}
The corpus underwent a multi-stage preprocessing procedure including text normalization, deduplication, length-based filtering, and the removal of entries with weak speculative relevance. Long-form source texts were segmented into retrieval-ready passages, while structurally different sources were reformatted into a unified text representation for downstream indexing.

Because the deployed system used retrieval rather than corpus-wide prompting, the full corpus was treated as an upstream candidate pool rather than a direct runtime resource. From this pool, \textbf{218 representative fragments} were manually retained as stylistic references for the generation pipeline. These fragments were selected for their relevance to speculative futures, technical imagination, and science-fictional narrative tone, and were organized into thematic groups to support retrieval during prompt construction.

\section{Physical Requirements and Demonstration Feasibility for \textit{Cosmic1001}}
\label{app:physical-requirements}

\subsection{Information Sheet}

\textbf{Exhibit type.}  
\textit{Cosmic1001} is a screen-based interactive installation consisting of one large display, one exhibition pedestal or stand, and a standard power connection.

\vspace{0.5em}
\noindent
\textbf{Display.}  
The work requires at least one large screen for visual presentation. A commercial display in the range of 55--65 inches is recommended. The display should support a minimum resolution of 1920 $\times$ 1080 and a refresh rate of at least 60\,Hz. Standard AC power input is sufficient for exhibition use.

\vspace{0.5em}
\noindent
\textbf{Example display specification.}
\begin{itemize}
    \item Recommended size: 55--65 inches
    \item Power: standard AC 100--240V
\end{itemize}

\vspace{0.5em}
\noindent
\textbf{Spatial layout.}  
The installation can be presented within a compact footprint of approximately 1.5\,m $\times$ 1.5\,m to 2\,m $\times$ 2\,m. The screen may be mounted on or positioned behind a pedestal, stand, or existing exhibition support. No enclosed structure is required.A minimum exhibition area of approximately 4\,m $\times$ 4\,m is recommended. The installation consists of one primary interaction station for user input and browsing, together with one or two additional display surfaces for presenting generated future news and the shared \textit{Future Tunnel} visualization. 

\vspace{0.5em}
\noindent
\textbf{Power requirements.}  
The installation requires only one standard power source, which may be connected through a regular extension cord or power strip. No specialized electrical infrastructure is needed.

\vspace{0.5em}
\noindent
\textbf{Lighting conditions.}  
Moderate or dim exhibition lighting is preferred to ensure clear screen visibility. Strong direct light on the display surface should be avoided.

\vspace{0.5em}
\noindent
\textbf{Sound environment.}  
If audio playback is enabled, a relatively quiet exhibition environment is preferred. If necessary, the work can also be presented without external speakers or adapted to headphone-based playback.

\vspace{0.5em}
\noindent
\textbf{Internet connection.}  
A stable Internet connection is recommended if the live AI generation pipeline is demonstrated. If network access is limited, the installation may also be adapted to a preloaded or partially offline presentation mode.

\subsection{Feasibility of Demonstration at the Exhibition Site}

\textit{Cosmic1001} is designed to be lightweight and technically compatible with standard exhibition conditions. The work does not require complex mounting, custom fabrication, or large-scale projection equipment. Its basic setup consists only of one commercial display, one pedestal or stand, and access to standard AC power.

The demonstration is therefore feasible in most gallery, museum, or conference exhibition environments that can accommodate a small screen-based installation. 
\end{document}